\documentclass[12pt,preprint,number,sort&compress,lontitle,times]{elsarticle}
\usepackage{graphicx}
\usepackage{epsfig}
\usepackage{lineno}
\usepackage{multirow}
\usepackage{fourier}
\usepackage{color}
\usepackage{appendix}
\usepackage[english]{babel}
\usepackage[]{geometry}
\usepackage{cleveref}
\newcommand{\dif}{\mathrm{d}}

\begin{document}

\begin{frontmatter}

\title{Coherent Suppression of Molecular Bremsstrahlung Radiation at GHz Frequencies in the Ionization Trail of Extensive Air Showers}

\author{Olivier Deligny}
\ead{deligny@ijclab.in2p3.fr}
\address{Laboratoire de Physique des 2 Infinis Ir\`ene Joliot-Curie, CNRS/IN2P3, Universit\'{e} Paris-Saclay, Orsay, France}

\begin{abstract}
Several attempts to detect extensive air showers (EAS) induced by ultrahigh-energy cosmic rays have been conducted in the last decade based on the molecular Bremsstrahlung radiation (MBR) at GHz frequencies from quasi-elastic collisions of ionisation electrons left in the atmosphere after the passage of the cascade of particles. These attempts have led to the detection of a handful of signals only, all of them forward-directed along the shower axis and hence suggestive of originating from geomagnetic and Askaryan emissions extending into GHz frequencies close to the Cherenkov angle. In this paper, the lack of detection of events is explained by the coherent suppression of the MBR in frequency ranges below the collision rate due to the destructive interferences impacting the emission amplitude of photons between the successive collisions of the electrons. The spectral intensity at the ground level is shown to be several orders of magnitude below the sensitivity of experimental setups. In particular, the spectral intensity at 10~km from the shower core for a vertical shower induced by a proton of $10^{17.5}$~eV is 7-to-8 orders of magnitude below the reference value anticipated from a scaling law converting a laboratory measurement to EAS expectations. Consequently, the MBR cannot be seen as the basis of a new detection technique of EAS for the next decades.
\end{abstract}
\end{frontmatter}
 

\section{Introduction}

The investigation and understanding of the intensity of cosmic rays with energies in excess of $10^{19}~$eV, particles discovered nearly 60 years ago~\cite{Linsley:1961kt,Linsley:1963km}, has been demanding for more and more precise data, both from the statistical and from the systematical point of view. Currently, the Pierre Auger Observatory, covering an area of 3000 km$^2$ in Argentina~\cite{ThePierreAuger:2015rma}, and the Telescope Array, covering an area of 700 km$^2$ -- planned to be extended to 3000 km$^2$ -- in the United States~\cite{AbuZayyad:2012kk}, are the two largest-ever built detectors of EAS induced by cosmic rays. A harvest of data is now allowing numerous constraints to be inferred on the acceleration mechanisms operating in the extragalactic astrophysical sites producing the particles, and on the energetics and the location of these sources~\cite{Aab:2020rhr,Aab:2018chp,Caccianiga:2020njq}. While the noose is tightening around some nearby extragalactic objects, no discrete source of ultrahigh-energy cosmic rays has been identified so far through an intense clustering of arrival directions. This does not preclude that sources may be captured on a collective basis in a near future, but another jump in statistics appears necessary. The pending challenge for the next generation of ground-based observatories is thus to provide this jump in statistics while preserving equal, or reaching better, performances in accuracy to measure the EAS characteristics. 

Historically, a breakthrough in the detection technique of EAS has been the use of fluorescence telescope stations that was pioneered first in tests at the Volcano Ranch experiment and then with the original Fly's Eye experiment~\cite{Bergeson:1977nw} made up of arrays of several hundred of photomultiplier tubes which, thanks to a set of telescope mirrors, each monitor a small portion of the sky. These sensors detect the fluorescence caused by the de-excitation of nitrogen molecules as a result of their excitation by the many ionising electrons created as the cascade passes through the atmosphere. This de-excitation gives rise to weak ultraviolet radiation, but which can be detected up to 30 or 40 km away on moonless nights and which offers the possibility to observe EAS side-on thanks to the isotropic emission. These telescopes thus allow a measurement of the longitudinal profile of the showers, which in turn is used to infer both the energy of the showers in a calorimetric way, without recourse to external information to calibrate the energy estimator, and the slant depth of maximum of shower development, a proxy, the best up to date, of the primary mass of the particles. Large detection areas can be covered by means of a few fluorescence stations only, spaced every 20 km or so. However, the flip side of the technique is its low duty cycle, about 10\%, due to the need for operating during moonless nights only. 
 
Through the passage of charged particles in the atmosphere, the energy of an EAS is deposited mainly by ionisation. The resulting numerous ionisation electrons can, in turn, produce their own emission such as continuum Bremsstrahlung emission through quasi-elastic scattering with molecular nitrogen and oxygen. Due to the expected isotropic and unpolarised emission, molecular Bremsstrahlung radiation (MBR) in the GHz band, which propagates in the atmosphere in a quasi-unattenuated way (less than 0.05~dB~km$^{-1}$), is thus providing a mechanism, with a 100\% duty cycle, for performing shower calorimetry in the same spirit as the fluorescence technique does, by mapping the ionisation content along the showers through the intensity of the microwave signals detected at ground level. 

Triggered by microwave emission measurements in laboratory~\cite{Gorham:2007af}, new telescope techniques based on the detection of the microwave emission in the GHz frequency range have been subsequently tested~\cite{ThePierreAuger:2013eja,Smida:2014sia,Gaior:2017rfa}. Only a few handful signals forward-directed along the shower axes were recorded, with in particular no side-on observation of EAS~\cite{ThePierreAuger:2013eja,Smida:2014sia}. Hence, the hopes raised a decade ago for a new breakthrough in the detection technique of EAS have been dashed, and the MBR technique does not even remain on the drawing board nowadays. The goal of this paper is to explain the reasons of the faintness of the emission that was not anticipated in previous estimates, which were based on the scaling of the radiation of a single quasi-elastic collision of a free electron with a neutral molecule to the total rate of quasi-elastic collisions in the short-lived electron/air plasma~\cite{Samarai:2014yda,Samarai:2016xpu}. 

Coherence effects for Bremsstrahlung processes in dense matter are well-known in the case of ultra-relativistic electrons undergoing multiple scattering on Coulomb centers: if an electron undergoes multiple scattering while traversing the ``formation zone'', the Bremsstrahlung amplitudes from before and after the scattering can interfere, reducing the probability of photon emission for photon energies below a certain value. This is the Landau-Pomeranchuk-Migdal effect~\cite{Landau:1953um,Migdal:1956tc}, which leads to a suppression for the Bremsstrahlung cross section compared to the Bethe-Heitler one of the single-scattering picture. In air, this suppression factor becomes important at GHz frequencies for electron kinetic energies greater than 1~MeV. In~\cref{sec:mbr}, effects of the same nature are shown to induce a suppression factor in the case of non-relativistic electrons and to be responsible, hence, for the faintness of the MBR emission of EAS in the GHz band. Once this suppression mechanism is established, it is straightforward to apply it to the short-lived electron/air plasma left after the passage of an EAS in~\cref{sec:plasma} and infer the expected spectral intensity from EAS at ground level as shown in~\cref{sec:shower}. A discussion of the results is given in~\cref{sec:discussion}.

\section{Molecular Bremsstrahlung Radiation of Low-Energy Electrons in a Dense Plasma}
\label{sec:mbr}

Let us consider an electron/neutral plasma with low-energy electrons colliding elastically with neutral molecules during a finite time and describe each electron as a classical charged particle coupled to a Maxwell field. In this framework, the energy radiated by the electron is associated with the deviations caused by the collisions with the neutral molecules: when an electron approaches a neutral molecule, the electric field of the electron polarises the neutral molecule, and this polarisation gives rise to a dipole moment that induces an attractive interaction potential at a short distance range. For non-relativistic particles, the spectral radiated energy per unit solid angle flowing into an elementary cone $\dif\Omega$ and received at distance $R$ assumed to be far away from the accelerated charge reads as
\begin{equation}
\label{eqn:Eomega}
    \mathcal{E}(\omega,\Omega)=\frac{e^2}{16\pi^3\epsilon_0c^3}\left\langle\left|\int\dif t'~\left(\mathbf{q}\times\left(\mathbf{q}\times\dot{\mathbf{v}}(t')\right)\right)\exp{(-i\omega t')}\right|^2\right\rangle,
\end{equation}
where $e$ is the electric charge, $\epsilon_0$ is the vacuum permittivity, $c$ is the speed of light, $\mathbf{q}$ is a unit vector in the observer direction that changes negligibly during a small acceleration interval and $\mathbf{v}(t')$ is the electron velocity at retarded time $t'$. The use of the $\langle\cdot\rangle$ symbol stands for the average over realisations of the stochastic process that governs the dynamics of $\mathbf{v}(t')$. An electron appearing free at $t'=0$ and disappearing (by attachment) at $t'=\tau$ experiences accelerations during each collision. To derive the expected radiation by accounting for the effect of successive collisions, the collisions are modeled as a random series of impulsive velocity changes $\Delta\mathbf{v}_k$ occurring during the finite time duration $\tau$. This implies that the acceleration of an electron can be written as
\begin{equation}
\label{eqn:dotv}
    \dot{\mathbf{v}}(t')=\sum_{k=1}^{N_{\mathrm{coll}}}\Delta\mathbf{v}_k\delta(t',t'_k),
\end{equation}
with $N_{\mathrm{coll}}$ a Poisson variable governed by the collision rate $\Gamma_{\mathrm{coll}}$. On inserting this expression into the angular frequency spectrum of the energy radiated by a non-relativistic accelerated particle, one gets, after integration over all directions:
\begin{equation}
\label{eqn:EomegaMC}
    \mathcal{E}(\omega)=\frac{e^2}{6\pi^2\epsilon_0c^3}\left\langle\left|\mathbf{v}_0+\sum_{k=1}^{N_{\mathrm{coll}}}\Delta\mathbf{v}_k\exp{(-i\omega t'_k)} -\mathbf{v}_{N_{\mathrm{coll}}}\exp{(-i\omega \tau)}\right|^2\right\rangle,
\end{equation}
where, to include the transition radiations associated with the appearance and disappearance of the electron, an effective acceleration has been introduced at the initial and final times, $\dot{\mathbf{v}}(0)=\mathbf{v}_0\delta(t',0)$ and $\dot{\mathbf{v}}(\tau)=-\mathbf{v}_{N_{\mathrm{coll}}}\delta(t',\tau)$. In contrast to the traditional recipe to derive $\mathcal{E}(\omega)$ for the Bremsstrahlung process consisting in multiplying the radiated energy of one single collision, $\mathcal{E}_1(\omega)$, by the number of collisions $N_{\mathrm{coll}}$, this expression accounts for coherence effects. For $\omega \gg \Gamma_{\mathrm{coll}}$, the random arguments in the exponential $\omega t_k$ are random numbers since the random times $t_k$ are of the order of $1/\Gamma_{\mathrm{coll}}$: the regime is then incoherent, and the scaling $\mathcal{E}(\omega)=N_{\mathrm{coll}}\mathcal{E}_1(\omega)$ holds. In contrast, for $\omega \ll \Gamma_{\mathrm{coll}}$, the random arguments are close to 0 so that all random phases are close to 1, and the radiation is then largely suppressed from the interference between photons emitted by different elements of electron pathlength. \\ 

The r.h.s. of \cref{eqn:EomegaMC} provides the relevant framework to simulate by Monte-Carlo the interferences that lead to the Bremsstrahlung suppression at angular frequencies smaller than the collision rate. However, a more explicit expression can be obtained starting from, similarly to the Landau-Pomeranchuk way, integrating by parts \cref{eqn:Eomega}. The antiderivative term is zero since $\mathbf{v}(0^-)=\mathbf{v}(\tau^+)=0$. After integration over all directions, one thus gets
\begin{equation}
\label{eqn:Eomega_bis}
    \mathcal{E}(\omega)=\frac{e^2\omega^2}{6\pi^2\epsilon_0c^3}\iint\dif t'\dif t''~\left\langle \mathbf{v}(t')\cdot\mathbf{v}(t'')\right\rangle\exp{(-i\omega (t'-t''))},
\end{equation}
so that the information is actually encompassed in the two-point correlation function of the electron velocity. The connection between \cref{eqn:EomegaMC} and \cref{eqn:Eomega_bis} is illustrated below with two scenarios for which both the Monte-Carlo and the analytical methods are used.  

\begin{figure}[!t]
\centering
\includegraphics[width=7cm]{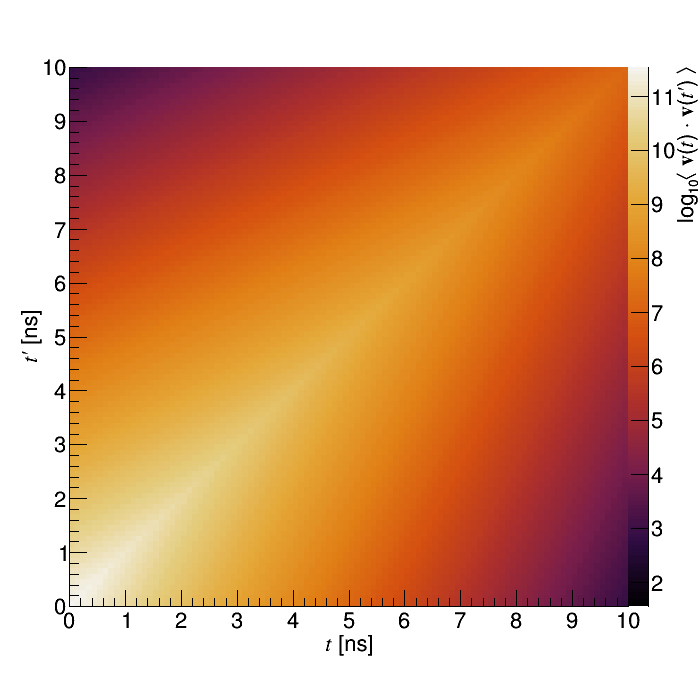}
\includegraphics[width=8cm]{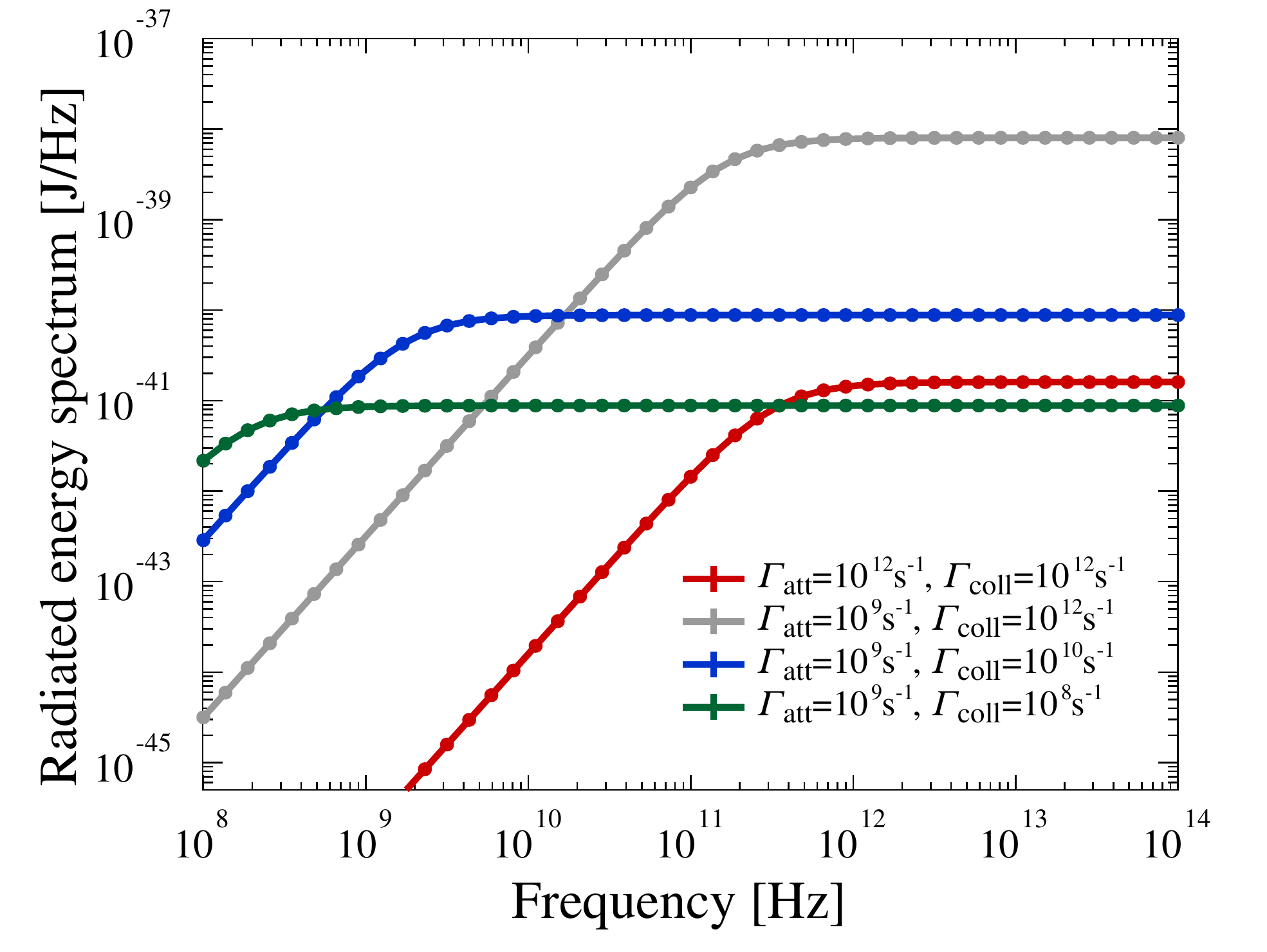}
\caption{\small{Left: Two-point correlation function of the electron velocity, obtained by Monte-Carlo, for electron kinetic energies of 1~eV, and $\Gamma_{\mathrm{att}}=\Gamma_{\mathrm{coll}}=10^{12}~$s$^{-1}$. Right: Spectrum of the radiated energy for several values of $\Gamma_{\mathrm{att}}$ and $\Gamma_{\mathrm{coll}}$.}}
\label{fig:vv1}
\end{figure}

First, let's consider the case of free electrons during a time duration $0\leq t\leq \tau$ undergoing elastic collisions at a rate $\Gamma_{\mathrm{coll}}$. The time duration $\tau$ is a random variable governed by a process of electron attachment at a rate $\Gamma_{\mathrm{att}}$, attachment process that does not need to be made explicit at this stage. Averaged over a large number of pathlengths of free electrons, the corresponding two-point correlation function of the electron velocity, obtained by Monte-Carlo, is shown in the left panel of \cref{fig:vv1} for electron kinetic energies of 1~eV, and $\Gamma_{\mathrm{att}}=\Gamma_{\mathrm{coll}}=10^{12}~$s$^{-1}$ so that the effect of both processes is made visible. The attachment process leads to a decorrelation of the velocities evolving as $\exp{\left(-\Gamma_{\mathrm{att}}~\mathrm{max}\left(t',t''\right)\right)}$. Besides, the spatial diffusion induced by the elastic collisions leads to a decorrelation that depends only on the time difference between $t'$ and $t''$, decorrelation hence evolving as $\exp{\left(-\Gamma_{\mathrm{coll}}\left|t'-t''\right|\right)}$. The two-point correlation function of the electron velocity thus reads as
\begin{equation}
\label{eqn:vv}
    \left\langle \mathbf{v}(t')\cdot\mathbf{v}(t'')\right\rangle=v_0^2\times\left\{
    \begin{array}{ll}
        \exp{\left(-\Gamma_{\mathrm{att}} t'-\Gamma_{\mathrm{coll}}\left(t'-t''\right)\right)} & \mbox{if } t'\geq t'', \\
        \exp{\left(-\Gamma_{\mathrm{att}} t''-\Gamma_{\mathrm{coll}}\left(t''-t'\right)\right)} & \mbox{otherwise.}
    \end{array}
\right.
\end{equation}
On inserting this expression into \cref{eqn:Eomega_bis}, the angular frequency spectrum of the radiated energy reads
\begin{equation}
\label{eqn:Eomega_ter}
    \mathcal{E}(\omega)=\frac{e^2v_0^2}{3\pi^2\epsilon_0c^3}\frac{\left(\Gamma_{\mathrm{att}}+\Gamma_{\mathrm{coll}}\right)\omega^2}{\Gamma_{\mathrm{att}}\left(\left(\Gamma_{\mathrm{att}}+\Gamma_{\mathrm{coll}}\right)^2+\omega^2\right)}.
\end{equation}
This expression exhibits the suppression of the emission for $\omega \ll \Gamma_{\mathrm{coll}}+\Gamma_{\mathrm{att}}$. On the other hand, for $\omega \gg \Gamma_{\mathrm{coll}}+\Gamma_{\mathrm{att}}$, the radiation scales as $(\Gamma_{\mathrm{coll}}/\Gamma_{\mathrm{att}})(e^2v_0^2/3\pi^2\epsilon_0c^3)\equiv N_{\mathrm{coll}}\mathcal{E}_1$ if $\Gamma_{\mathrm{coll}} \gg \Gamma_{\mathrm{att}}$, and as $\mathcal{E}_1$ if $\Gamma_{\mathrm{coll}} \ll \Gamma_{\mathrm{att}}$, as expected. These features are illustrated in the right panel of \cref{fig:vv1}, where the spectrum of the radiated energy is shown as a function of the frequency $\nu=\omega/2\pi$ for several values of $\Gamma_{\mathrm{att}}$ and $\Gamma_{\mathrm{coll}}$. Overlaid on the points obtained from the Monte-Carlo computation by means of \cref{eqn:EomegaMC}, the continuous line obtained from \cref{eqn:Eomega_ter} is in perfect agreement. How the spectral suppression and the overall scale of the radiation are governed by these two parameters is clearly visible.

\begin{figure}[!t]
\centering
\includegraphics[width=7cm]{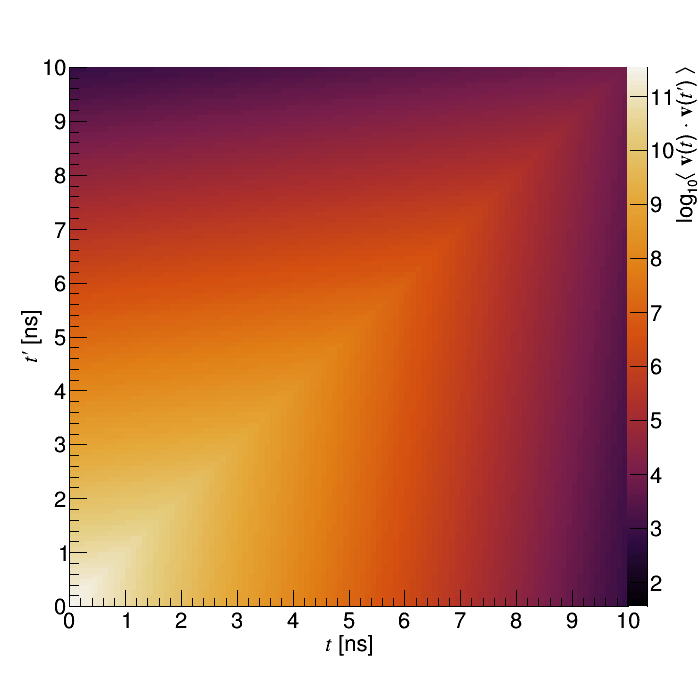}
\includegraphics[width=8cm]{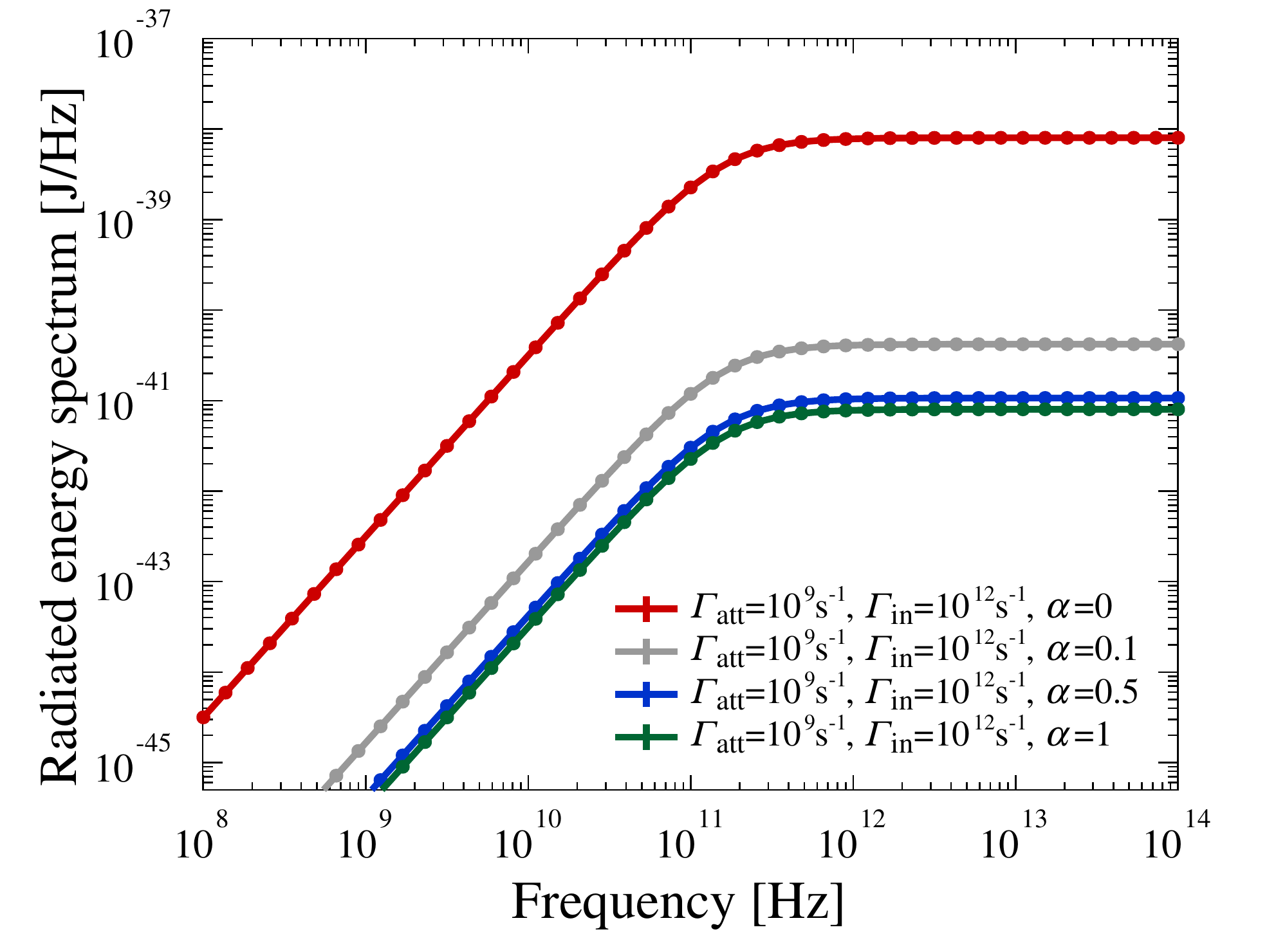}
\caption{\small{Left: Two-point correlation function of the electron velocity, obtained by Monte-Carlo, for electron kinetic energies of 1~eV, $\Gamma_{\mathrm{att}}=\Gamma_{\mathrm{in}}=10^{12}~$s$^{-1}$, $\Gamma_{\mathrm{el}}=0$ and $\alpha=0.5$. Right: Spectrum of the radiated energy for $\Gamma_{\mathrm{att}}=\Gamma_{\mathrm{in}}=10^{12}~$s$^{-1}$ and several values of $\alpha$.}}
\label{fig:vv2}
\end{figure}

More complex is the case considering inelastic collisions in addition to elastic ones. For clarity, the corresponding rates are denoted as $\Gamma_{\mathrm{el}}$ (elastic) and $\Gamma_{\mathrm{in}}$ (inelastic). The electron is supposed to loose a constant fraction of velocity, $\alpha v$, at every inelastic collision. This scenario is not necessarily of relevant interest in practice, but it provides us with an analytical solution that allows a direct understanding of the impact of the inelastic processes on the radiated energy. The result of the Monte-Carlo for the $\left\langle \mathbf{v}(t')\cdot\mathbf{v}(t'')\right\rangle$ function is shown in the left panel of \cref{fig:vv2} for, as previously, electron kinetic energies of 1~eV, $\Gamma_{\mathrm{el}}=0$, $\Gamma_{\mathrm{att}}=\Gamma_{\mathrm{in}}=10^{12}~$s$^{-1}$, and for $\alpha=0.5$. The velocity losses induce a faster net decorrelation than in the case of elastic collisions: in addition to the spatial diffusion term depending on the time difference $|t'-t''|$, an additional evolving term $\exp{\left(-\alpha\left(2-\alpha\right)\Gamma_{\mathrm{in}}~\mathrm{min}\left(t',t''\right)\right)}$ is observed so that the expression sought for the two-point correlation function of the electron velocity reads as
\begin{equation}
\label{eqn:vv_bis}
    \left\langle \mathbf{v}(t')\cdot\mathbf{v}(t'')\right\rangle=v_0^2\times\left\{
    \begin{array}{ll}
        \exp{\left(-\Gamma_{\mathrm{att}}t'-\left(\Gamma_{\mathrm{el}}+\Gamma_{\mathrm{in}}\right)\left(t'-t''\right)-\alpha\left(2-\alpha\right)\Gamma_{\mathrm{in}}t''\right)}& \mbox{if } t'\geq t'', \\
        \exp{\left(-\Gamma_{\mathrm{att}}t''-\left(\Gamma_{\mathrm{el}}+\Gamma_{\mathrm{in}}\right)\left(t''-t'\right)-\alpha\left(2-\alpha\right)\Gamma_{\mathrm{in}}t'\right)}& \mbox{otherwise.}
    \end{array}
\right.
\end{equation}
The angular frequency spectrum of the radiated energy reads in this case
\begin{equation}
\label{eqn:Eomega_qua}
    \mathcal{E}(\omega)=\frac{e^2v_0^2}{3\pi^2\epsilon_0c^3}\frac{\left(\Gamma_{\mathrm{att}}+\Gamma_{\mathrm{el}}+\Gamma_{\mathrm{in}}\right)\omega^2}{\left(\Gamma_{\mathrm{att}}+\alpha(2-\alpha)\Gamma_{\mathrm{in}}\right)\left(\left(\Gamma_{\mathrm{att}}+\Gamma_{\mathrm{el}}+\Gamma_{\mathrm{in}}\right)^2+\omega^2\right)},
\end{equation}
which highlights, through the term $\Gamma_{\mathrm{att}}+\alpha(2-\alpha)\Gamma_{\mathrm{in}}$ in the denominator, the additional suppression of the emission induced by the inelastic process. This is illustrated in the right panel of \cref{fig:vv2} for $\Gamma_{\mathrm{att}}=\Gamma_{\mathrm{in}}=10^{12}~$s$^{-1}$, $\Gamma_{\mathrm{el}}=0$ and several values of $\alpha$. For $\alpha=0$, the impact of the inelastic collisions is identical to that of elastic ones, while for $\alpha=1$, the impact is that of an attachment at a rate $\Gamma_{\mathrm{in}}$.  \\

The classical electrodynamics approach used here to derive the Bremsstrahlung emission is justified by the weakness of the photon energies in the considered frequency range. From a quantum perspective, the production of photons with energies $h\nu$, with $h$ the Planck constant, corresponds to transitions between unquantised energy states of the free electrons (``free-free'' transitions). In the framework of non-equilibrium quantum field theory, it can be shown that the quasi-free scattering approximation indeed breaks down due to the successive collisions through a careful classification of diagrams and an appropriate re-summation of subsets of graphs~\cite{Knoll:1995nz}. The suppression factors obtained in this framework are however depending on the relaxation rate of the source, which is less straightforward to infer than the collision rates used in the approach adopted here.

\section{Emission from the Ionisation Trail Left After the Passage of an Extensive Air Shower}
\label{sec:plasma}

The energy of an EAS is, as already stressed, deposited mainly through the ionisation process through the development of the cascade in the atmosphere. Let $n_{\mathrm{EAS}}$ be the number of high-energy charged particles per surface unit in the cascade and $\rho(\mathbf{x})$ the density of molecular nitrogen or oxygen in the atmosphere at the position $\mathbf{x}$. These high-energy electrons/positrons from the cascade are refereed to as ``primary electrons'' hereafter, in contrast to the ionisation electrons, the production of which per unit volume, per velocity band and per time unit follows from 
\begin{equation}
\label{eqn:nei}
n(\mathbf{x},\mathbf{v}_0,t_0)=\frac{\rho(\mathbf{x})f(\mathbf{v}_0,t_0)}{I_0+\left\langle T\right\rangle}~\left\langle\frac{\dif E}{\dif X}\right\rangle~n_{\mathrm{EAS}}(\mathbf{x}).
\end{equation}
Here, $I_0$ is the ionisation potential to create an electron-ion pair in air, the bracketed expression $\left\langle\dif E/\dif X\right\rangle$ stands for the mean energy loss of the EAS charged particles per grammage unit, and $f(\mathbf{v}_0,t_0)$ is the distribution in velocity and time of the resulting ionisation electrons, which is related to that expressed in terms of kinetic energy, $f_T\left(T\right)$, through the Jacobian transformation
\begin{equation}
\label{eqn:f}
f(\mathbf{v}_0,t_0)=\frac{mv_0}{4\pi\left(1-\left(v_0/c\right)^2\right)^{3/2}}f_T\left(T\left(v_0\right),t_0\right),
\end{equation}
with $m$ the mass of the electron. For primary charged particles in the cascade with $\geq~$MeV energies, an expression for the distribution $f_T\left(T,t_0=0\right)$ that accounts for relativistic effects as well as indistinguishability between primary and secondary electrons, which modify the low-energy behaviour~\cite{Rosado:2014bya}, follows from that provided in~\cite{Arqueros:2009zz}:
\begin{equation}
\label{eqn:f0}
f_T(T)=\frac{8\pi ZR_y^2}{m\left(\beta(T_{\mathrm{p}})c\right)^2}\frac{1+C\exp{(-T/T_\mathrm{k})}}{T^2+\overline{T}^2},
\end{equation}
where $R_y$ is the Rydberg constant, $\beta(T_{\mathrm{p}})$ is the relativistic factor for the primary electron with energy $T_{\mathrm{p}}$, $T$ ranges from 0 to $T_{\mathrm{max}}=(T_{\mathrm{p}}-I_0)/2$ due to the indistinguishability between primary and secondary electrons, the constant $C$ is determined in the same way as in~\cite{Opal:1971} so that $\int \mathrm{d}T~f_T(T)$ reproduces the total ionisation cross section, $T_\mathrm{k}=77~$eV is a parameter acting as the boundary between close and distant collisions, and $\overline{T}$ is a measured parameter such that $\overline{T}=13.0~(17.4)~$eV for nitrogen (oxygen). In the energy range of interest, this expression leads to $\left\langle T\right\rangle\simeq 40~$eV, in agreement with the well-known stopping power. The remaining time dependence in $t_0$, reflecting the subsequent cascade of ionisation electrons produced by secondary electrons themselves as long as their kinetic energy is above $I_0$, is derived by Monte-Carlo below.

As long as they remain free, ionisation electrons with density $n'\equiv n(\mathbf{x}',\mathbf{v}'_0,t'_0)$ can thus produce photons through the process of quasi-elastic collisions with neutral molecules in the atmosphere with an angular frequency spectrum 
\begin{equation}
\label{eqn:Eomega_final}
    \mathcal{E}(\omega)=\frac{e^2\omega^2}{6\pi^2\epsilon_0c^3}\iint\dif \mathbf{x'}\dif \mathbf{x''}\iint\dif \mathbf{v}'_0\dif \mathbf{v}''_0\iint\dif t'_0\dif t''_0\int_{t'_0}^{\infty}\int_{t''_0}^{\infty}\dif t'\dif t''~\left\langle \left(n'\mathbf{v}(t')\right)\cdot\left(n''\mathbf{v}(t'')\right)\right\rangle e^{-i\omega (t'-t'')},
\end{equation}
where, compared to the single-particle case presented in \cref{sec:mbr}, the two-point correlation function of the electron velocities must now account for the density of particles. For an incoherent process between independent particles such as the MBR, this two-point correlation function is diagonal in every variable governing the densities:
\begin{equation}
\label{eqn:2pt}
    \left\langle \left(n'\mathbf{v}(t')\right)\cdot\left(n''\mathbf{v}(t'')\right)\right\rangle=n'\delta(\mathbf{x}',\mathbf{x}'')\delta(\mathbf{v}'_0,\mathbf{v}''_0)\delta(t'_0,t''_0)~\left\langle \mathbf{v}(t')\cdot\mathbf{v}(t'')\right\rangle. 
\end{equation}
In this way, the radiation scales with the number of particles once integrating \cref{eqn:Eomega_final} over positions, initial velocities and initial creation time. 

\begin{figure}[!t]
\centering
\includegraphics[width=10cm]{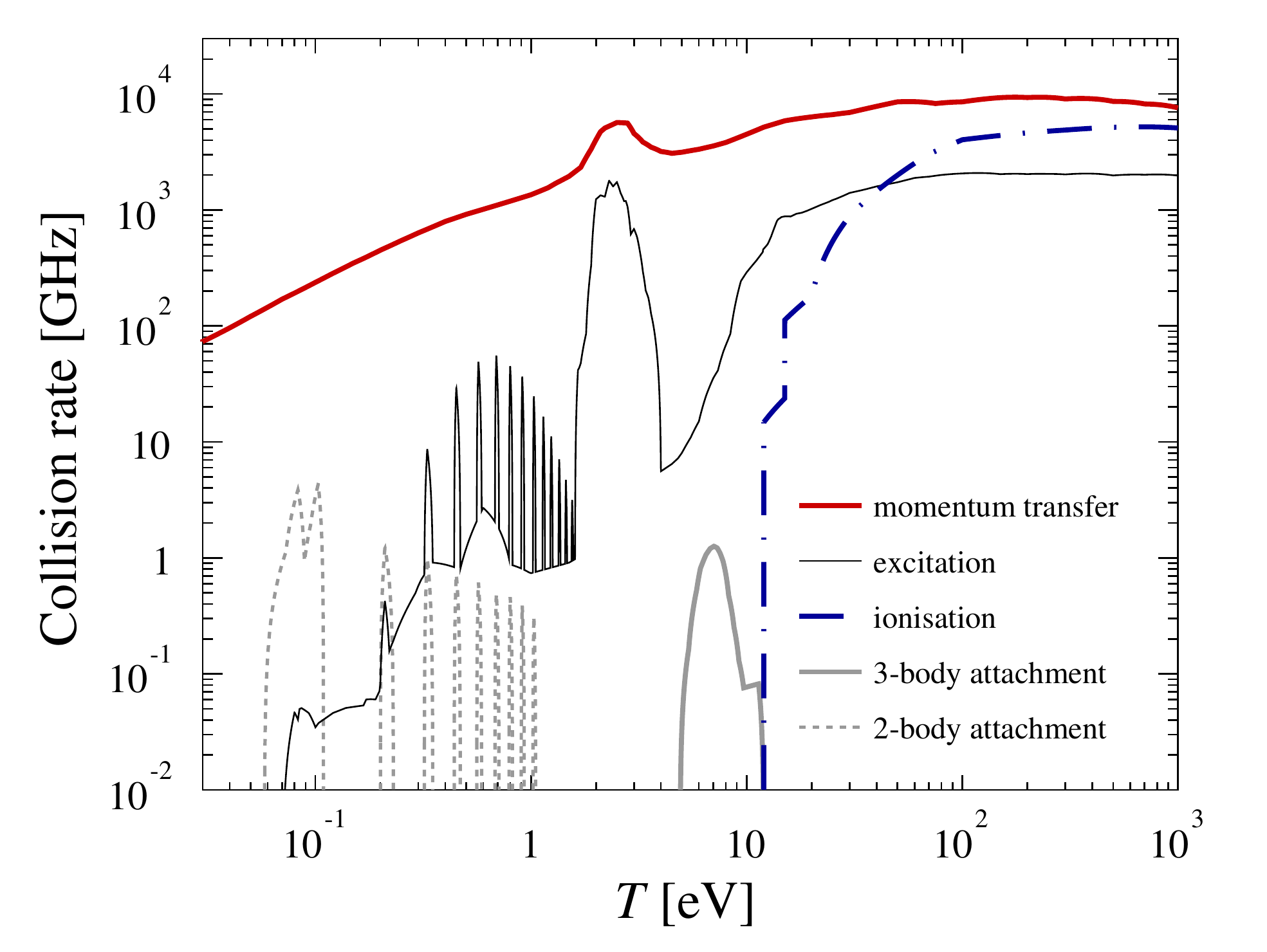}
\caption{\small{Collision rates of free electrons in air as a function of their kinetic energy.}}
\label{fig:rates}
\end{figure}

The radiation is thus determined, as in the simple examples in \cref{sec:mbr}, by the two-point correlation function of the velocities of a single electron, obtained by Monte-Carlo by simulating a large number of test particles with initial velocities drawn at random from \cref{eqn:f} and undergoing collisions, the rate of which being taken from experimental tabulated data in~\cite{JILA}. The main features of the different rates, shown in \cref{fig:rates}, depend on the energy. The total momentum transfer collision rate goes from $\simeq 100~$GHz up to $\simeq 10~$THz in the explored kinetic energy range, with different inelastic contributions depicted by the different curves. Ionisation on N$_2$ and O$_2$ molecules dominates the collisions for $T\geq 40~$eV, causing energy losses on a time scale below the picosecond. Excitation on electronic levels of N$_2$ and O$_2$ molecules enters into play in a dominant way below 40~eV down to 4~eV, with energy losses that occur on time scales going from picoseconds to a few nanoseconds when going down in energy. Below 4~eV down to 1.7~eV, resonances for excitation on N$_2$ and O$_2$ molecules through ro-vibrational processes cause energy losses on a time scale of the picosecond. Then, below 1.7~eV down to 0.2~eV, resonances for excitation on N$_2$ and O$_2$ molecules through ro-vibrational processes and for two-body attachment process on O$_2$ molecules enter into play. These processes are quantised in energies. The energy losses of the excitation resonances occur on a time scale of a few tens of picoseconds, while the time scale of disappearance of the electrons through the two-body attachment process is of the order of the nanosecond. Despite their low abundance, CO$_2$ and H$_2$O molecules induce energy losses on a time scale of a nanosecond that degrade electron energies down to 0.1~eV, where the two-body attachment process make them disappearing on a time scale of a few nanoseconds. Excitations of H$_2$O molecules are also considered, the concentration of which is subject to large variations in the atmosphere; a typical value of 3,000~ppm is used in this study. Below $T=0.1~$eV, the contribution of the electrons to the total radiation is negligible and is not considered. 

\begin{figure}[!t]
\centering
\includegraphics[width=10cm]{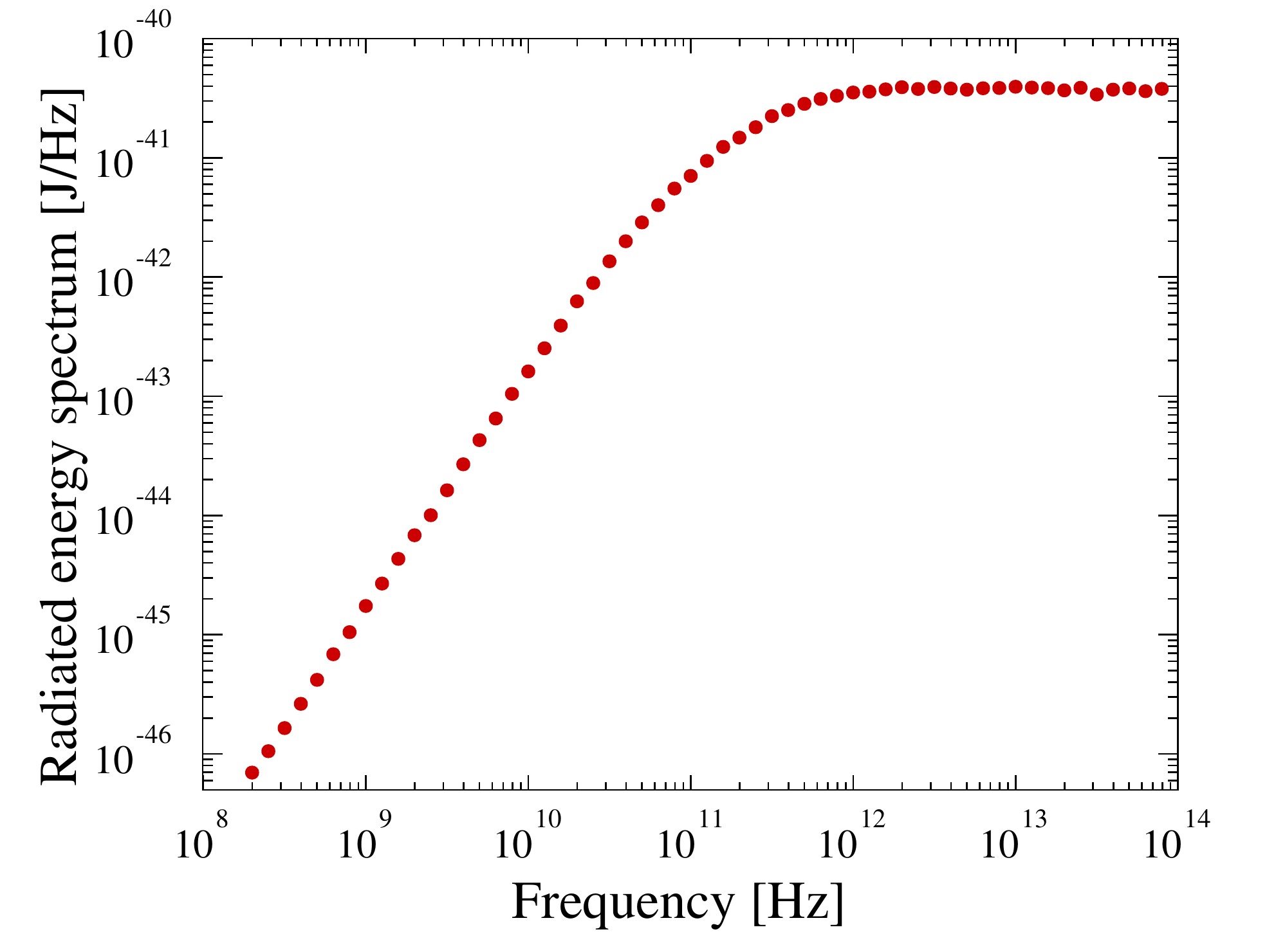}
\caption{\small{Frequency spectrum of the radiated energy by the ionisation trail of an EAS.}}
\label{fig:rad}
\end{figure}

The tabulated two-point correlation function of the velocities of a single electron obtained from the Monte-Carlo simulation allows the determination of the spectrum of radiation sought for. Note that in the simulation, each newly produced ionisation electron is stacked and subsequently simulated so as to add its contribution, as well as that of all possible ``daughter'' particles, to the radiation of the electron initially tracked. This is equivalent to considering $f(\mathbf{v}_0,t_0)=f(\mathbf{v}_0)\delta(t_0,0)$, which is the relevant quantity after carrying out the changes of variables $t'\rightarrow t'-t'_0$ and $t''\rightarrow t''-t''_0$ in \cref{eqn:Eomega_final}. The resulting spectrum of radiation from the ionisation trail left after the passage of an EAS, normalised to the contribution of one single particle, is shown in \cref{fig:rad}. The radiation is observed to be suppressed below THz frequencies, with a quadratic dependence in frequency in the GHz range.

\section{Spectral Intensity at GHz Frequencies Expected from Extensive Air Showers}
\label{sec:shower}

\begin{figure}[!t]
\centering
\includegraphics[width=8cm]{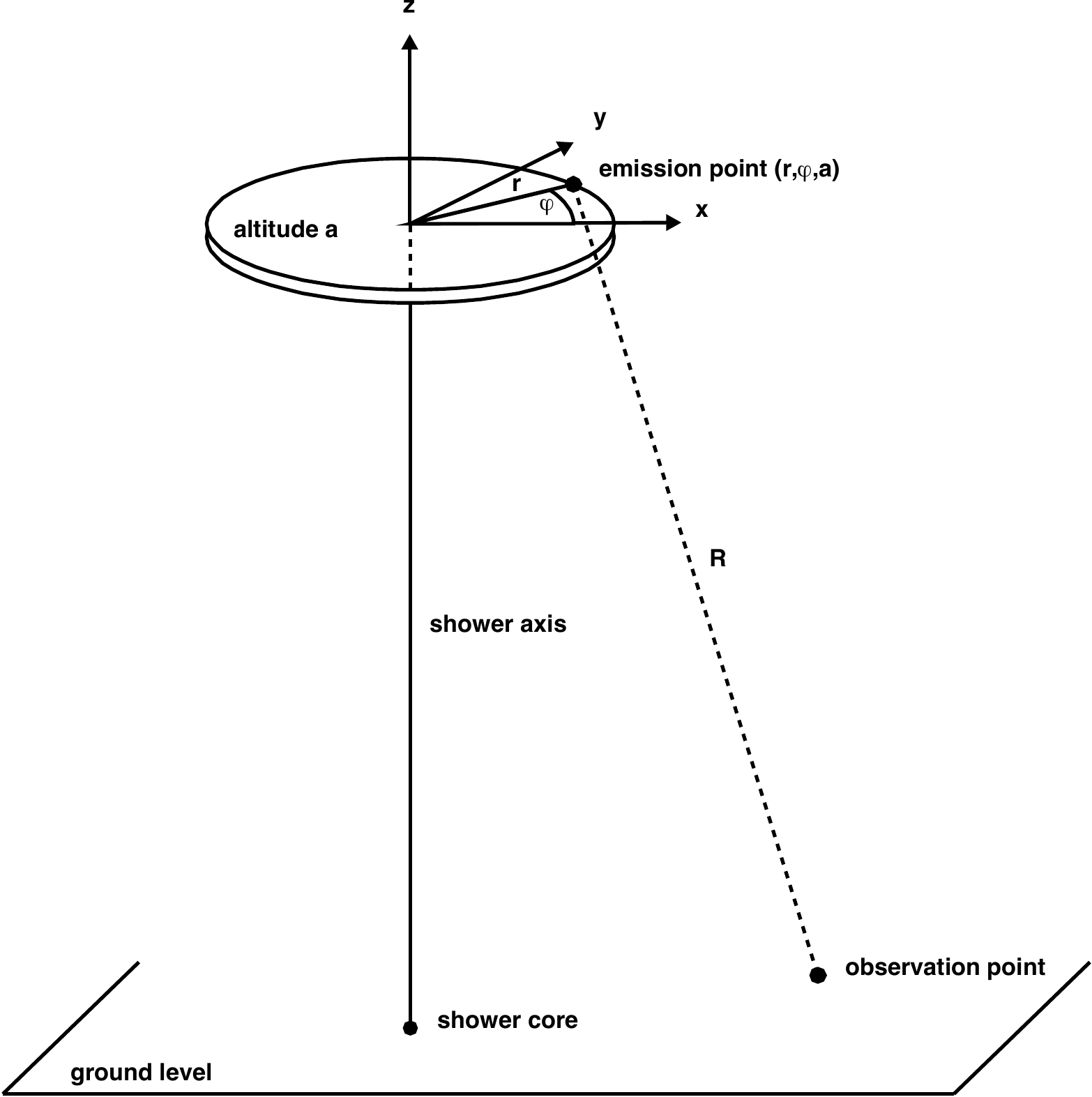}
\caption{\small{Geometry of a vertical EAS used throughout the paper.}}
\label{fig:EAS}
\end{figure}

The function that describes the power of the radiation received per unit frequency and passing through any unit area at an observation point $\mathbf{x}_{\mathrm{obs}}$ is the spectral intensity, $\Phi(\nu,\mathbf{x}_{\mathrm{obs}})$. It results from the summation of the radiation emitted by all the ionisation electrons produced along the shower track. Expressed in W~m$^{-2}$~Hz$^{-1}$ units, it is the quantity directly accessible to the experiment in a frequency band. To provide relevant orders of magnitude for the spectral intensities that can be expected from MBR in the GHz band at the ground level, a crude model of EAS, limited to a vertical incidence, is used to infer an expression of $n_{\mathrm{EAS}}(\mathbf{x})$ to be plugged into \cref{eqn:nei}, the geometry of which is depicted in \cref{fig:EAS}. The spectral intensity then results from 
\begin{equation}
\label{eqn:Phi}
\Phi(\nu,\mathbf{x}_{\mathrm{obs}})=\iiint \frac{r\dif r\dif\varphi\dif a}{4\pi R^2(r,\varphi,a)}~\mathcal{P}(\nu,a),
\end{equation}
where $\mathcal{P}(\nu,a)$ is the frequency spectrum of emitted power, obtained by dividing the radiated energy by the mean duration of the emission identified as the ``mean lifetime of the plasma'', which amounts, from the simulations described in the previous section, to $\simeq 30~$ns at the ground level.

Following~\cite{Samarai:2014yda}, the EAS is considered as a thin disk of high-energy charged particles propagating in the atmosphere at the speed $c$.  The density of particles in the disk depends on the distance $r$ to the axis. Restricting ourselves to the electromagnetic component, which is the dominant component producing ionisation electrons, the lateral extension of the cascade can be expressed in terms of the Moli\`ere radius $R_M$, which is such that 90\% of the energy is contained within this distance from the axis. In this way, the number of electrons/positrons per surface unit $n_{\mathrm{EAS}}$ at any position $\mathbf{x}=(r,\varphi,a)$ is known to be well reproduced by the NKG profile~\cite{Greisen:1956,1958PThPS...6...93K}:
\begin{equation}
\label{eqn:neas}
n_{\mathrm{EAS}}(\mathbf{x})=N(a)\times C(s(a))~R_M^{-2}~\left(\frac{r}{R_M}\right)^{s(a)-2}\left(1+\frac{r}{R_M}\right)^{s(a)-4.5}.
\end{equation}
Here, $s(a)$ stands for the age parameter at altitude $a$ defined as $s(a)=3X(a)/(X(a)+2X_{\mathrm{max}})$,
and $C(s)$ is a normalisation factor such that $2\pi\int r\dif r~n_{\mathrm{EAS}}=N(a)$, where $N(a)$ is the number of electrons/positrons at any altitude $a$. For a given primary type and a given energy $E$, this latter quantity follows from the Gaisser-Hillas parameterisation of the longitudinal development of the electromagnetic cascade, which depends only on the cumulated slant depth $X$ expressed as the ratio between the vertical thickness of the atmosphere $X_{\mathrm{vert}}$ ($\sim~$1000 g~cm$^{-2}$ at sea level) and the cosine of the zenith angle of the EAS~\cite{1977ICRC....8..353G}:
\begin{equation}
N(a)=N_{\mathrm{max}}\bigg(\frac{X(a)-X_0}{X_{\mathrm{max}}-X_0}\bigg)^{\frac{X_{\mathrm{max}}-X_0}{\lambda}}\exp{\bigg(\frac{X_{\mathrm{max}}-X(a)}{\lambda}\bigg)},
\end{equation}
with $X(a)$ the depth corresponding to the altitude $a$, $X_0$ the depth of the first interaction,
$X_{\mathrm{max}}$ the depth of shower maximum, $N_{\mathrm{max}}$ the number of particles
observed at $X_{\mathrm{max}}$, and $\lambda$ a parameter describing the attenuation of the shower.

\begin{figure}[!t]
\centering
\includegraphics[width=10cm]{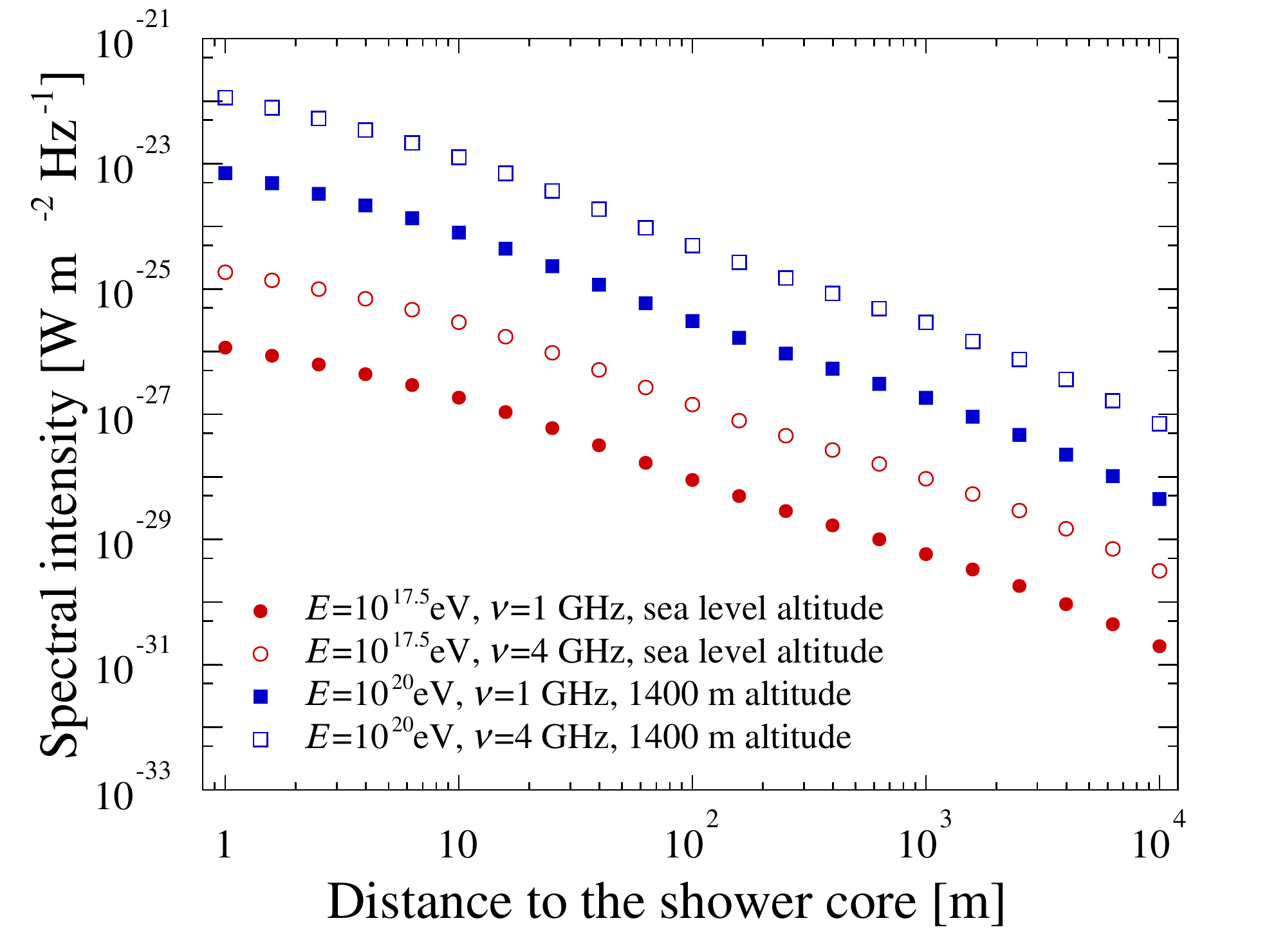}
\caption{\small{Spectral intensity as a function of the distance to the shower core.}}
\label{fig:spectral_intensity}
\end{figure}

The spectral intensity expected at different distances from the shower core is shown in \cref{fig:spectral_intensity}, for two primary energies and two frequencies of interest. The quadratic dependence in frequency is seen. The rapid decrease in amplitude for increasing distances is striking. Published limits on the MBR emission are currently at the level of $10^{-14.5}$~W~m$^{-2}$~Hz$^{-1}$~\cite{Alvarez_Mu_iz_2012}. The values derived in this study are by orders of magnitude below these current limits. A relevant estimate of the minimal spectral intensity $\Phi_{\mathrm{min}}$ detectable by an antenna operating in a bandwidth $\Delta\nu$ with a noise temperature $T_{\mathrm{sys}}$ and an effective area $A_{\mathrm{eff}}$ is known to obey
\begin{equation}
\label{eqn:Phimin}
\Phi_{\mathrm{min}}=\frac{kT_{\mathrm{sys}}}{A_{\mathrm{eff}}\sqrt{\tau\Delta\nu}},
\end{equation}
where $k$ is the Boltzmann constant and $\tau$ the receiver sampling time. For values $\Delta\nu=0.8$~GHz, $\tau=10~$ns, $T_{\mathrm{sys}}=50~$K and $A_{\mathrm{eff}}$ approaching $10^3~$cm$^2$, values typical of the setups used at the Pierre Auger Observatory, for instance, one gets $\Phi_{\mathrm{min}}$ on the order of a few $10^{-21}$~W~m$ ^{-2}$~Hz$^{-1}$. Based on previous estimates of MBR spectral intensities, such a sensitivity was anticipated to allow the detection of high-energy showers within a kilometer from the core~\cite{Samarai:2014yda,Samarai:2016xpu}. By contrast, the results obtained in this study show that the expected signals are out of reach of the experimental setups, even for a $10^{20}$~eV shower sampled at 1400~m altitude level, that of the Pierre Auger Observatory.  The spectral intensities are 7-to-8 orders of magnitude below the reference values anticipated from a scaling law converting the laboratory measurement to EAS expectations put forward in~\cite{Gorham:2007af} and 5-to-6 orders of magnitude below the values estimated in~\cite{Samarai:2014yda,Samarai:2016xpu}.

\section{Discussion}
\label{sec:discussion}

The coherent suppression of the MBR in the GHz frequency range as described in \cref{sec:mbr} is thus prohibitive to allow experimental setups using antennas to detect EAS crossing the field of view of the receivers. This coherent suppression stems from the destructive interferences impacting the emission amplitude of photons between the successive collisions of the same electron. The spectral intensity at the ground level is several orders of magnitude below the sensitivity of experimental setups. The few detected events over the past years in this frequency range cannot be due to MBR from the ionisation electrons left along the shower track. Other radio-emission mechanisms, such as the geomagnetic effect, the Askaryan effect or the MBR from the primary electrons of the showers, are likely responsible for the observed forward-directed signals. No side-on observation of EAS is, however, expected from these emission mechanisms, which consequently cannot be seen as the basis of a new breakthrough in the detection technique of EAS for the next decades.

For frequencies above the collision rate, the contribution of the MBR to the air-fluorescence yield, $Y$, estimated in~\cite{Samarai:2016zhk} is also affected by the treatment of the successive collisions presented in \cref{sec:mbr} due to the impact of the inelastic collisions that quench the emission compared to the simple scaling $\mathcal{E}=N_{\mathrm{coll}}\mathcal{E}_1$. Considering $n_l=\rho f(\mathbf{v}_0) \langle\dif E/\dif X\rangle/(I_0+\langle T\rangle)$ as the number of ionisation electrons per length and velocity units, the number of emitted photons is then estimated by plugging $n_l$ into \cref{eqn:2pt} and \cref{eqn:Eomega_final}, normalised by $h\nu$. The MBR contribution to $Y$ is then obtained by normalising the number of emitted photons to the deposited energy per unit length and by integrating over the frequency range corresponding to the UV [330-400] nm wavelength range:
\begin{equation}
Y=\frac{2\pi}{I_0+\left\langle T\right\rangle}\int\frac{\dif\nu}{h\nu}\int\dif\mathbf{v}_0f(\mathbf{v}_0)\iint\dif t'\dif t''~\left\langle \mathbf{v}(t')\cdot\mathbf{v}(t'')\right\rangle e^{-i2\pi\nu (t'-t'')}.
\end{equation}
This yields to $Y\simeq 10^{-4}~$MeV$^{-1}$, which is three orders of magnitude lower than the estimate provided in~\cite{Samarai:2016zhk}. The contribution of the MBR to the fluorescence yield, the world average value from various experiments is $(7.04 \pm 0.24)$MeV$^{-1}$~\cite{Rosado:2014bya}, is thus negligible. 

Finally, the effect of the successive collisions on the radiation of the ionisation electrons, as presented in this work, is important to study the possible radar echoes of cascades of particles, which have been recently measured at SLAC in ice to a level that may lead to a viable neutrino detection technology for energies above $10^{16}~$eV~\cite{Prohira:2019glh}. The adaptation of the formalism to account for the incoming wave in the two-point correlation function of the electron velocities and for the coherence of the re-radiation in \cref{eqn:2pt} can be used to quantify the contribution of the re-radiation of the incoming wave by the ionisation electrons~\cite{Deligny2021}.

\section*{Acknowledgements}

I thank Carola Dobrigkeit, Antoine Letessier-Selvon, Lorenzo Perrone and Frank Schroeder for their feedback and careful reading of the paper. This work benefited from the support of the French Agence Nationale de la Recherche (ANR) under reference ANR-12-BS05-0005-01 at an earlier stage. 

\section*{References}

\bibliographystyle{elsarticle-num}
\bibliography{biblio}

\end{document}